\documentstyle[preprint,aps]{revtex}

\begin{document}
\draft
\tighten
\parindent 0cm

\title{MUONIC MOLECULES OF CHARGE Z$\ge$3: COULOMBIC PROPERTIES AND
NUCLEAR TRANSITIONS}

\author{V. B. Belyaev$^1$,  M. Decker$^2$, H. Fiedeldey$^3$, \\
S. A. Rakityansky$^1$, W. Sandhas$^2$, S. A. Sofianos$^3$}
\address{
$^1$Joint Institute  for Nuclear Research, 141980 Dubna, Russia}
\address{
$^2$ Physikalisches Institut, Universit\"at Bonn, D-53115 Bonn, Germany}
\address{
$^3$ Physics Department, University of South Africa, P.O.Box 392,
Pretoria 0001, South Africa}
\date{\today}
\maketitle

%%%%%%%%%%%%%%%%%%%%%%%%%%%%%%%%%%%%%%%%%%%%%%%%%%%%%%%%%%%%%%%%%%%%%%%%
\vspace{1cm}
\centerline{Invited talk: International Workshop on Muon Catalyzed
Fusion, Krakov, 1993.}
\centerline{To appear in {\bf Nucleonica}}
\vspace{2cm}
%\centerline{{\bf VERSION 3.0}}
%\vspace{2cm}
%%%%%%%%%%%%%%%%%%%%%%%%%%%%%%%%%%%%%%%%%%%%%%%%%%%%%%%%%%%%%%%%%%%%%%%%

\begin{abstract}
In this report we will discuss properties of and models for nuclear
transitions in muonic molecules formed via collisions of muonic atoms
of hydrogen isotopes with light nuclei like $L\!i$ and $B\!e$. Their
importance for nuclear astrophysics and nuclear physics at super low
energies is emphasized.
\end{abstract}

%%%%%%%%%%%%%%%%%%%%%%%%%%%%%%%%%%%%%%%%%%%%%%%%%%%%%%%%%%%%%%%%%%%%%%%%
\section{Introduction}

In this report we will mainly consider three particle systems of the
type $h\,L\!i\,\mu$ and $h\,B\!e\,\mu$, where $h$= p, d, or t are
isotopes of hydrogen, $L\!i =\,^6\!L\!i\,$ or $^7\!L\!i$, and
$B\!e =\,^7\!B\!e,$ $^8\!B\!e$ or $^{10}\!B\!e$. There are strong
indications that these particles can form molecular three-body
resonances which, however, have not yet been observed experimentally.
One of these indications is that $d\,H\!e\,\mu$ resonances were seen in
collisions of $p\mu$ and $d\mu$ atoms with $^3\!H\!e$ and $^4\!H\!e$
\cite{bys}.
In the $L\!i$- or $B\!e$-case there is, of course, a stronger
repulsion (proportional to $Z$, the charge of the heavy nucleus) which,
however, is partly compensated by an attractive force (proportional to
$Z^2$) due to the polarizibility of the $h\mu$ atoms. Theoretical
estimations \cite{ari,bel1,ger}, in fact, show that $h\,L\!i\,\mu$
and $h\,B\!e\,\mu$ resonances do exist. It is however not only the
existence, but in particular the formation probability of such
systems, which is to be investigated.

Up to now, only mechanisms connected with the transfer of energy to
the electronic degrees of freedom in the $L\!i$ or $B\!e$ atom
(Auger transitions \cite{kra1,bel2} or resonance excitations of
atoms in the final state \cite{kra2}) have been considered in the
literature. However, it is to be expected that some other, also
resonant, mechanisms exist, leading to the excitation of a
(in the Born-Oppenheimer sense) 'slow' degree of freedom. That is,
the analog of the Vesman mechanism \cite{ves} for $dt\mu$ formation
should play an essential role also in the present case.
Indeed, $L\!i$ atoms are usually not free in a hydrogen medium, but
can form molecular hydrides of $L\!i$, like $L\!iH$, or $L\!i_2$
molecules.  The whole spectrum of excitations of such molecules is
displaced into the region of a few $eV$. That means, if in the
three-particle systems $h\,L\!i\,\mu$ states exist with a few $eV$
energy, a resonant formation of the Vesman type is possible. This
observation explains, why it is so important to perform precise
calculations of the spectra of muonic hydrides of $L\!i$ with the
different isotopes of hydrogen and $L\!i$. We shall return to this
problem later on and discuss the possibility of a realization of
the cycling process in a mixture of deuterium and $L\!i$.

If the $h\,L\!i\,\mu$ or $h\,B\!e\,\mu$ resonances are formed,
nuclear transitions can occur in them. Generally, these transitions
are expected to be suppressed in comparison with the analogous ones
in the $h\,^3\!H\!e\,\mu$ or $h\,^4\!H\!e\,\mu$ molecules, simply
due to their larger size. However, one can find a few peculiar
cases, where the nuclear reactions are of 'long range' in the
nuclear scale of distances. There are two reasons for the occurrence
of such long-range strong interactions. The first one is
related to the presence of a nuclear resonance in the compound systems,
close to the threshold energy of some two-body channels and sometimes
even in coincidence with it. In other words, long-range nuclear
transitions in the molecular system can be expected, when the heavy
constituents of a molecule would have a nuclear resonance at zero
relative energy. In this case we expect a rather large value for the
overlap integral, which characterizes the probability of a transition
between the molecular and the nondecreasing nuclear resonance wave
functions. As an example of such a situation, we consider the molecule
$d\,^7\!L\!i\,\mu$. In this case, the resonance state of the
$^9\!B\!e(5/2^+)$ nucleus occurs near the threshold energy of
the two-body channel $d\,+\,^7\!L\!i$.
The second reason for long-range nuclear transitions is connected
with the occurrence of two closely spaced threshold energies for some
two-body channels. It is almost evident that, if transitions between
such two channels take place, a small momentum transfer would be
transmitted in the transition process, and this implies a long range
of the interaction.

As an example of such a situation we consider the threshold energies
of the two-body states $d+\,^7\!B\!e$ and $p+\,^8\!B\!e^*(2^+)$ or
$p+\,^{10}\!B\!e$ and $t+\,^8\!B\!e(g.s.)$. In the first example the
difference between the threshold energies of the systems $d+\,^7\!B\!e$
and $p+\,^8\!B\!e^*(2^+)$ approximately equals $0.04 MeV$. In the
second example it is equal to $0.004 MeV$. Below, we will estimate
the probabilities of nuclear transitions in the systems
$d\,^7\!B\!e\,\mu$ and $p\,^{10}\!B\!e\,\mu$.

These systems are of interest also from another points of view. There
is a considerable lack of information about the nuclear interactions
in the $eV-keV$ energy region. Such energies, however, are attainable
in muonic systems. By lack of information, we mean not only that the
dynamics of nuclear transitions in this energy interval is not well
understood, but also that the validity of the usual properties of the
strong interactions is not guaranteed. Properties of strong
interactions as charge symmetry, iso-invariance, the character of
P- and T-invariance (or its violation) are all established mainly in
the $MeV$ region and, up to now, have only been extrapolated to the
low-energy region.

A further reason for the interest in such molecules
is related to questions of nuclear astrophysics.
Here we mean nuclear reactions which took place in the process of
primordial nucleosynthesis just after the Big Bang, and those
occuring in those stars, where light elements are produced.
One of the problems here is the abundance of $^7\!L\!i$ in the
Universe. In order to discuss this problem reliably, we first
establish that the abundance of $^7\!L\!i$ seen in the Universe is
indeed due to primordial nucleosynthesis. Evidence for this
conjecture was found some years ago from the observation of the
abundance of $^7\!L\!i$ in old stars \cite{mau}.

%\vspace{1cm}
%\centerline{\fbox{\rule[-2.5cm]{0cm}{5cm}This is a quite nice place for Figure
%%1}}
%\centerline{
%{\bf Fig. 1:} The correlation between the abundance of $L\!i$ and $F\!e$
%              in old stars}
%\centerline{(normalized to the abundance in the sun).}
%\vspace{1cm}

In Figure 1 the abundance of $^7\!L\!i$ is shown as a function of the
abundance of $F\!e$. It can be compared with the abundance of $M\!g$
(see Figure 2), an element which was definitely not created in
primordial nucleosynthesis, but in the stars.

%\vspace{1cm}
%\centerline{\fbox{\rule[-2.5cm]{0cm}{5cm}This is a quite nice place for Figure
%%2}}
%\centerline{
%{\bf Fig. 2:} The correlation between the abundance of $M\!g$ and $F\!e$
%              in old stars}
%\centerline{(normalized to the abundance in the sun).}
%\vspace{1cm}

This comparison indeed indicates that the abundance of $^7\!L\!i$
nuclei, in contrast to the case of $M\!g$, is constant and therefore
cannot be attributed to the processes occuring in the stars. The modern
theory \cite{wag} of the creation of light elements during the
primordial nucleosynthesis, predicts the following distributions as
functions of the nucleon-to-photon ratio $\eta$.

%\vspace{1cm}
%\centerline{\fbox{\rule[-2.5cm]{0cm}{5cm}This is a quite nice place for Figure
%%3}}
%{\bf Fig. 3:} $Y_P$ denotes the predicted primordial abundance of
%$^4\!H\!e$ relative to $H$ as a function of $\eta$ for three different
%numbers of neutrino species $N_\nu$. The other curves show the
%relative abundances of $D$, $^3\!H\!e$, and $^7\!L\!i$ as a function
%of $\eta$.
%\vspace{1cm}

In this theory one has to use the cross sections for the nuclear
reactions in the $keV$ range of energies. These energies are not
accessible in laboratory experiments, when we want to study
for instance reactions involving $^7\!L\!i$ nuclei, such as
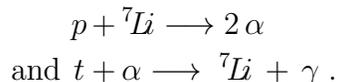
\begin{eqnarray}
\nonumber
&p&\,+\,^7\!L\!i \longrightarrow 2\,\alpha\\
\nonumber \mbox{ and }
&t&\,+\, \alpha  \longrightarrow \;^7\!L\!i\, + \, \gamma \; .
\end{eqnarray}
We can try to get access to these data via the the observation of such
transitions in the muonic systems $p\,^7\!L\!i\,\mu$ and
$t\,^4\!H\!e\,\mu$
$$
p\;^7\!L\!i\,\mu \longrightarrow \left \{ \begin{array}{ll}
& ^8\!B\!e\,\mu +\gamma \\
&  ^8\!B\!e +\mu
\end{array}\right.
$$
and
$$
t\; ^4\!H\!e\,\mu \longrightarrow \left \{ \begin{array}{ll}
& ^7\!L\!i\,\mu +\gamma \\
&  ^7\!L\!i +\mu
\end{array}\right. \; ,
$$
where the relative energy of the heavy particles in the molecule is
arround a few $eV$ only.

Usually, astrophysicists extrapolate the astrophysical
S-factor\footnote{The S-factor is assumed to be a slow function
of the energy, if no resonances in the region concerned are present.},
defined as
\begin{equation}
\sigma=\frac{e^{-2\pi\eta(E)}}{E}S(E) \; ,
\end{equation}
into the region of low energies. How dangerous such a process is, can be
seen from the example of the extrapolation procedure for the S-factor
of the reaction
\begin{equation}
d+d\longrightarrow \;^4\!H\!e+\gamma \; .
\end{equation}
In Figure 4 we show the extrapolated S-factor and the S-factor, fitted
to the new data at low energies \cite{rol}. It is seen that the
difference in the 'old' and 'new' S(0)-values is two orders of
magnitude.

%\vspace{1cm}
%\centerline{\fbox{\rule[-2.5cm]{0cm}{5cm}This is a quite nice place for Figure
%%4}}
%{\bf Fig. 4:} New experimentall data (at energies less than $100 keV$)
%and old data (above $100 keV$) for the S-factor. The full calculation
%(solid line) is shown together with a calculation using only the
%($L=2$,$S=2$) configuration (short-dashed line).
%{\em The other two curves? (MD)}
%\vspace{1cm}

Another reaction that we discuss in the context of muonic systems
is the radiative capture of protons by the $^7\!B\!e$ nucleus
\begin{equation}
p + \; ^7\!B\!e \longrightarrow \; ^8\!B + \gamma \; .
\end{equation}
It is well known \cite{bah} that this transition is the last step
in the pp-chain in the sun  and is the bottleneck, which is crucial
for the determination of the flux of high energy neutrinos from the
sun \cite{jon}. In laboratory measurements of the cross section of this
process only the region of energies down to $100 keV$ is accessible.
An extrapolation to the energies at the centre of the sun ($1.5 keV$)
has to be made. Here again one of the possibilities to obtain the
value of the S-factor in the few $eV$ energy region
is to search for this transition in the muonic molecule
$p\,^7\!B\!e\,\mu$.
In this case we do not have a threshold resonance in the $p+\,^7\!B\!e$
system, but a very peculiar final state. The wave function of the
ground state of the $^8\!B$ nucleus has a very long tail due to the
(on a nuclear scale) extremely small separation energy of the proton
in $^8\!B$, which equals to $0.14 MeV$. Due to this long tail, we again
have a process, which occurs at large distances and can be favourable
for its observation in the muonic system $p\,^7\!B\!e\,\mu$.

%%%%%%%%%%%%%%%%%%%%%%%%%%%%%%%%%%%%%%%%%%%%%%%%%%%%%%%%%%%%%%%%%%%%%%%%
\section{Theoretical description of charge nonsymmetric muonic
molecules}

In this chapter we describe the treatment of the three-body
Coulomb problem $hZ\mu$. Since we have two heavy particles and one
light particle, it seems that the Born-Oppenheimer approach will be
reasonable to describe this system. Indeed, some estimations
\cite{ari} of the properties of the $h\,H\!e\,\mu$ systems performed by
this method (in the one-channel approximation) provided a
qualitatively correct description of some properties. Of course, the
quantitative description requires improvements since, within this
approximation, it is impossible to reproduce even the spectra of
the usual muonic  molecules of hydrogen. Moreover, since we are
interested in the probabilities of some partial transitions, say
the probability of formation, it will be more reliable to use a
method taking angular momentum conservation into account. Another
problem within the framework of the Born-Oppenheimer approach is that
it is very difficult to treat the contribution from the continuum
spectra of the two-center problem correctly. Thus, to avoid these
difficulties we will use another adiabatic formulation, namely the
hyperspherical adiabatic approach \cite{mac,bel3}. This method is
based on an expansion of the three-body wave function into the
so-called surface functions \cite{kup}.

Let us start the description of our formalism regarding Figure 5, where
the Jacobi coordinates are shown.

%\vspace{1cm}
%\centerline{\fbox{\rule[-1cm]{0cm}{2cm}
%\unitlength=0.5mm
%\begin{picture}(110,110)
%\put(0,0){\vector(1,2){39}}
%\put(10,20){\vector(3,1){89}}
%\put(18,47){\makebox(0,0)[br]{$\vec r$}}
%\put(60,33){\makebox(0,0)[tl]{$\vec R$}}
%\put(0,0){\circle*{3}}
%\put(40,80){\circle*{3}}
%\put(100,50){\circle*{3}}
%\put(110,52){\makebox(0,0)[t]{$h$}}
%\put(0,-5){\makebox(0,0)[t]{$Z$}}
%\put(45,95){\makebox(0,0)[t]{$\mu^-$}}
%\end{picture}
%}}
%\centerline{{\bf Fig. 5:} Jacobi coordinates for the system $hZ\mu$.
%\vspace{1cm}}

In these variables the Hamilton operator of the three-body
Coulomb system takes the form
\begin{equation}
H = - \frac{1}{2m} \Delta_{\vec r} - \frac{1}{2M} \Delta_{\vec R}
- \frac{Z e^2}{r} - \frac{e^2}{|\vec R - \beta \vec r|}
+ \frac{e^2}{|\vec R + \gamma \vec r|} \; ,
\end{equation}
where
\begin{equation}
m^{-1} = m_\mu^{-1} + m_Z^{-1}, \;\;\;
M^{-1} = m_h^{-1} + (m_\mu + m_Z)^{-1}, \;\;\;
\beta = \frac{m_Z}{m_\mu + m_Z}, \;\;\;
\gamma = 1 - \beta
\end{equation}
and $m_\mu$, $m_h$ and $m_Z$ denote the masses of the muon, the
hydrogen isotop and the heavier nucleus (with charge $Z$), respectively.
Introducing dimensionless variables according to
\begin{equation}
\vec x = \frac{\vec r}{a}, \;\;\; \vec y = \frac{\vec R}{\alpha a} \; ,
\end{equation}
with $\alpha=\sqrt{\frac{m}{M}}$ and $a$ being the Bohr radius of the
$(Z\mu)$ subsystem, $a=(Zme^2)^{-1}$, the Hamiltonian reads
\begin{equation}
H = \frac{1}{2 m a^2}
\Bigg \{
- \Delta_{\vec x} - \Delta_{\vec y} - \frac{2}{x}
- \frac{2}{Z |\alpha \vec y - \beta \vec x|}
+ \frac{2}{Z |\alpha \vec y + \gamma \vec x|}
\Bigg \} \; .
\label{hamil}
\end{equation}
This suggests to take $(2ma^2)^{-1}$ as our energy unit. Let us
define the hyperradius $\rho$ and the hyperangle $\omega$
via
\begin{equation}
x = \rho \cos \omega, \;\; y = \rho \sin \omega \; ,
\end{equation}
with $0 \leq \rho < \infty$ and $0 \leq \omega \leq \pi/2$. The
Hamiltonian (\ref{hamil}) then takes the well-known form
\begin{equation}
H = - \rho^{-5/2} \frac{\partial^2}{\partial \rho^2} \rho^{5/2}
+ \frac{\Lambda^2(\Omega)}{\rho^2} + \frac{{\cal V}(\Omega)}{\rho}\;,
\end{equation}
where the grand angular momentum operator $\Lambda(\Omega)$ is given
by
\begin{equation}
\Lambda^2(\Omega) = - \frac{1}{\cos \omega \sin \omega}
\frac{\partial^2}{\partial \omega^2} \cos \omega \sin \omega
+ \frac{{\vec l}^2_{\hat x}}{\cos^2 \omega}
+ \frac{{\vec l}^2_{\hat y}}{\sin^2 \omega} - \frac{1}{4} \; .
\end{equation}
$\vec l_{\hat x}$ and $\vec l_{\hat y}$ are the orbital angular
momentum operators corresponding to the variables $\hat x = \vec x/x$
and $\hat y = \vec y/y$, respectively. For the potential we have
\begin{equation}
{\cal V}(\Omega) = -2 \Bigg \{ \frac{1}{\cos \omega}
+ \frac{1}{Z|\alpha \hat y \sin \omega - \beta \hat x \cos \omega|}
- \frac{1}{Z|\alpha \hat y \sin \omega + \gamma \hat x \cos \omega|}
\Bigg \} \; .
\end{equation}
The five angles symbolized by $\Omega = (\omega,\hat x,\hat y)$ together
with the hyperradius $\rho$ provide a complete set of variables to
describe the positions of all three particles. The Schr\"odinger
equation therefore is
\begin{equation}
\Bigg \{
- \rho^{-5/2} \frac{\partial^2}{\partial \rho^2} \rho^{5/2}
+ \frac{\Lambda^2(\Omega)}{\rho^2} + \frac{{\cal V}(\Omega)}{\rho}
\Bigg \} \; \Psi(\rho,\Omega) = E \; \Psi(\rho,\Omega) \; .
\label{schroe}
\end{equation}

Let us now introduce the surface functions mentioned above. We want
to use these functions as a basis for the solutions $\Psi(\rho,\Omega)$
of (\ref{schroe}). For this purpose we consider the operator containing
only the angular part of the Hamiltonian,
\begin{equation}
\Bigg \{
\frac{\Lambda^2(\Omega)}{\rho^2} + \frac{{\cal V}(\Omega)}{\rho}
\Bigg \} \; B_n(\rho,\Omega) = U_n(\rho) \; B_n(\rho,\Omega) \; .
\label{adia}
\end{equation}
Its eigenfunctions $B_n(\rho,\Omega)$ are the so-called {\em surface
functions}, its eigenvalues $U_n(\rho)$ are called {\em
eigenpotentials}. In equation (\ref{adia}) the hyperradius $\rho$
enters only as a parameter. Since the surface functions form a
complete set on the
sphere of constant values of the hyperradius, the three-body
wave function can be expanded according to
\begin{equation}
\Psi(\rho,\Omega) = \rho^{-5/2} \sum_n f_n(\rho) B_n(\rho,\Omega) \; .
\label{psi}
\end{equation}
The adiabatic approximation consists in the truncation of this sum
into a finite number of terms. In contrast to the Born-Oppenheimer
approach, not the distance between the two heavy particles, but the
hyperradius (a measure of the average size of the whole system)
is the 'slow' variable. Both variables coincide if the
mass of the light particle (here $m_\mu$) tends to zero. The two
major advantages of our approach, in comparison with the
Born-Oppenheimer one, are that here the total angular momentum is a
good quantum number and that the eigenpotentials $U_n(\rho)$ tend to
the exact two-body binding energies in the limit of
$\rho \longrightarrow \infty$. Both properties can be derived from
equation (\ref{adia}).

Instead of solving the coupled system of differential equations, which
is obtained by inserting (\ref{psi}) into the Schr\"odinger equation
(\ref{schroe}), we here only treat the so-called {\em extreme
adiabatic approximation}, which is given by the neglect of the
derivatives of the surface functions with respect to the hyperradius.
The remaining radial equation then is
\begin{equation}
f_n''(\rho) = [ U_n(\rho) - E ] f_n(\rho) \; ,
\label{eaa}
\end{equation}
looking exactly as a radial two-body Schr\"odinger equation, where
$U_n(\rho)$ acts as an effective potential. Hence the name
'eigenpotential'.

The main task now consists in the determination of these
eigenpotentials. As it can be seen from equation (\ref{adia}),
for small values of the hyperradius, $\Lambda^2/\rho^2$ dominates over
the Coulomb potential ${\cal V}/\rho$. Therefore, the surface functions
in this region should be proportional to the eigenfunctions of
$\Lambda^2$, which are the well-known \cite{fab} hyperspherical
harmonics
$Y_{[\cal L]}(\Omega)$
\begin{equation}
Y_{[\cal L]}(\Omega) = N_{[\cal L]} \;
\cos^{l_x}\omega \; \sin^{l_y}\omega
\; P_k^{(l_y+\frac{1}{2},l_x+\frac{1}{2})}(\cos 2\omega) \;
{\cal Y}_{l_x l_y}^{LM}(\hat x,\hat y) \; .
\label{hh}
\end{equation}
These functions are characterized by the set of quantum numbers
$[{\cal L}] = \{k,l_x,l_y,L,M\}$. They contain the Jacobi polynomial
\begin{equation}
P_k^{(\alpha,\beta)}(x) =
\frac{(x-1)^{- \alpha} (x+1)^{- \beta}}{2^k k!}
\frac{d^k}{d x^k} \left [ (x-1)^{k+\alpha} (x+1)^{k+\beta} \right ]
\end{equation}
and the bispherical harmonics (eigenfunctions of the squared total
angular momentum operator ${\vec L}^2$)
\begin{equation}
{\cal Y}_{l_x l_y}^{LM}(\hat x,\hat y) = \sum_{m_x m_y}
{\langle\,l_x m_x l_y m_y \,|\,L M \,\rangle}
\; Y_{l_x m_x}(\hat x)
\; Y_{l_y m_y}(\hat y)
\end{equation}
with the Clebsch-Gordan coefficients $
{\langle\,l_x m_x l_y m_y \,|\,L M \,\rangle}
$.
The normalization constant $N_{[\cal L]}$ is given by
\begin{equation}
N_{[\cal L]} = \sqrt \frac
{2k! \; (l_x+l_y+2k+2) \; \Gamma(l_x+l_y+k+2)}
{\Gamma(l_x+k+\frac{3}{2}) \; \Gamma(l_y+k+\frac{3}{2})} \; .
\end{equation}

Large values of the hyperradius and negative energies (the only
energies we are interested in) correspond to the physical situation
where the muon is bound by one of the positive charges. Hence, the
surface functions in this region look like channel functions \cite{lin}
\begin{equation}
\Phi_{[m_i]}(\rho,\Omega) = \rho^{3/2} \;
{\cal R}_{nl_x}(\rho \cos \omega) \;
\sin^{l_y} \omega \;
{\cal Y}_{l_x l_y}^{LM}(\hat x,\hat y) \; .
\label{cf}
\end{equation}
Here, ${\cal R}_{nl_x}(x)$ denotes a hydrogen-like wave function and
$[m]$ the set of quantum numbers $\{n,l_x,l_y,L,M\}$. The index 'i'
specifies by which of the two nuclei the muon is bound.

To represent the surface functions in the whole space of $\rho$ we
use the following ansatz
\begin{equation}
B_n(\rho,\Omega) =
\sum_{i=1}^2 \sum_{[m_i]} \;
a_{n[m_i]}(\rho) \;
\Phi_{[m_i]}(\rho,\Omega) \;
+ \; \sum_{[\cal L]} \;
b_{n[\cal L]}(\rho) \;
Y_{[\cal L]}(\Omega) \; ,
\end{equation}
which (inserted in equation (\ref{adia})) yields a generalized
eigenvalue problem for the determination of the coefficients
$a_{n[m_i]}(\rho)$ and $b_{n[\cal L]}(\rho)$ and the eigenpotentials
$U_n(\rho)$.

Within this framework we treated the systems $dt\mu$,
$h\,^6\!L\!i\,\mu$, $h\,^7\!L\!i\,\mu$ and $h\,^7\!B\!e\,\mu$ in the
states with the total angular momentum $L=0$.
In all calculations we use 120 hyperspherical functions (\ref{hh}).
In the case of ($dt\mu$) we take 6 channel functions (\ref{cf}) and for
the systems of higher charge we take 10 channel functions into account.
The latter is useful due to the higher polarizibility of the $(h\mu)$
atom by the nucleus of higher charge. Tables I to IV show the binding
energies of the various muonic molecules using the extreme adiabatic
approximation (\ref{eaa}) in comparison with the results of References
\cite{kra1,mon}.

%%%%%%%%%%%%%%%%%%%%%%%%%%%%%%%%%%%%%%%%%%%%%%%%%%%%%%%%%%%%%%%%%%%%%%%%
\section{Estimations of nuclear transition probabilities in muonic
molecules of \mbox{$Be$} isotopes}

We start by considering simpler systems with respect to their nuclear
structure, e.g. the system $d\,^7\!B\!e\,\mu$, to demonstrate the
approximations we make and to discuss their reliability \cite{bel4}.
First, consider the structure of the spectrum of the $^8\!B\!e$ nucleus
shown in Figure 6.

%\vspace{1cm}
%\centerline{\fbox{\rule[-2.5cm]{0cm}{5cm}Insert Figure 6 (called Figure 7 in
%%the FAX)}}
%\centerline{
%{\bf Fig. 6:} The spectrum of the $^8\!B\!e$ nucleus (schematically)
%\cite{ajz1}.}
%\vspace{1cm}

As one can immediately see that is the case of two thresholds close
to each other, which correspond to the states $d+\,^7\!B\!e$ and
$p+\,^8\!B\!e(2^+)$. The difference between these threshold energies
is equal to $0.044 MeV$, a very small value on the nuclear scale.
In such a way, if the nuclear transition
$d+\,^7\!B\!e \longrightarrow p+\,^8\!B\!e(2^+)$
takes place, only a small momentum transfer is possible and large
distances between $d$ and $^7\!B\!e$ are involved.

Now, let us formulate the description of such a transition in the muonic
system
\begin{equation}
d\,^7\!B\!e\,\mu \longrightarrow p\,^8\!B\!e(2^+)\, \mu \; .
\label{dbm}
\end{equation}
We will treat both states (initial and final) as four-body systems
$n+p+\,^7\!B\!e+\mu$. The internal structure of the $^7\!B\!e$ nucleus
is not important in the range of energy and momentum transfer involved.
For that reason we will treat the excited state of the $^8\!B\!e(2^+)$
nucleus in a simple two-body $n+\,^7\!B\!e$ model.
Let us introduce the two sets of Jacobi coordinates, shown in Figure
7, which are appropriate for the description of the initial and final
states.

%\vspace{1cm}
%\centerline{\fbox{\rule[-2.5cm]{0cm}{5cm}Insert Figure 7 (called Figure 8 in
%%the FAX)}}
%\centerline{
%{\bf Fig. 7:} Two different sets of Jacobi coordinates appropriate to
%describe}
%\centerline{the initial and the final states of the four-body system.}
%\vspace{1cm}

The total Hamiltonian can be written in the form
\begin{equation}
H = H_0 + V^C + V^S \; ,
\label{fullh}
\end{equation}
where the sum of the kinetic energies $H_0$ is given by
\begin{equation}
H_0 = h_0(\vec r) + h_0(\vec \rho) + h_0(\vec R) \; ,
\end{equation}
and the potentials of the Coulomb, $V^C$, and strong, $V^S$, interactions
are
\begin{equation}
V^C = V^C_{^7\!B\!e\mu}
    + V^C_{p\mu}
    + V^C_{p\;^7\!B\!e}
\mbox{\hspace{1.5cm}and\hspace{1.5cm}}
V^S = V^S_{np}
    + V^S_{n\;^7\!B\!e}
    + V^S_{p\;^7\!B\!e} \; ,
\end{equation}
respectively. To continue, it is useful to introduce the auxiliary
channel Hamiltonians $H_{1,2}$, defined via
\begin{eqnarray}
H_1 = & & h_0(\vec r_1) + h_0(\vec \rho_1) + h_0(\vec R_1)
+ V^S_{np}(\vec r_1) \nonumber \\ & &
+ V^C_{^7\!B\!e\mu}
+ V^C_{d\mu}
+ V^C_{d\;^7\!B\!e}
\label{h1}
\end{eqnarray}
and
\begin{eqnarray}
H_2 = & & h_0(\vec r_2) + h_0(\vec \rho_2) + h_0(\vec R_2)
+ V^S_{n\;^7\!B\!e}(\vec r_2) \nonumber \\ & &
+ V^C_{^8\!B\!e\mu}
+ V^C_{p\mu}
+ V^C_{p\;^8\!B\!e} \; ,
\label{h2}
\end{eqnarray}
and the corresponding Schr\"odinger equations for the channel
eigenfunction\footnote{For reasons of simplicity we will neglect for
the moment the contributions from the non-point-like charge
distributions in the $d$ and $^8\!B\!e$ nuclei.},
\begin{equation}
H_i \psi_i = \varepsilon_i \psi_i \; ,
\mbox{\hspace{1.5cm}} i=1,2 \; .
\end{equation}
As one can see from the equations (\ref{h1}) and (\ref{h2}) the channel
Hamiltonians do not contain the strong interaction between the clusters
$d$ and $^7\!B\!e$ in the initial state and between $p$ and $^8\!B\!e$
in the final one. Therefore they describe the three-body Coulomb systems
$d\,^7\!B\!e\,\mu$ and $p\,^8\!B\!e\,\mu$ and the internal motions in
the deuteron and $^8\!B\!e$ nucleus, and therefore are still
four-body Hamiltonians.

To solve the Schr\"odinger equation
\begin{equation}
H \Psi = E \Psi
\label{schr}
\end{equation}
with the full Hamiltonian (\ref{fullh}) we use the ansatz
\begin{equation}
\Psi = c_1 \psi_1 + c_2 \psi_2 \; .
\label{lcao}
\end{equation}
This ansatz reminds us of the well-known LCAO (Linear Combination
of Atomic Orbitals) approximation, where the role of the
'atomic orbitals'
is played by the nuclear states of $d$ and $^8\!B\!e(2^+)$. The
approximation (\ref{lcao}) therefore implies that the internal motion
in the deuteron does not disturb the internal motion in the $^8\!B\!e$
nucleus much and vice versa. Let us discuss the reliability of that
ansatz for the solution of equation (\ref{schr}). We should first
note that in the usual molecular ion of the hydrogen $H_2^+$, where
the LCAO approximation is qualitatively a good one, the size of the
atomic orbital has the same order of magnitude as the size of the
$H_2^+$ ion itself, namely $r_0 \sim 10^{-8} cm$. In the case of the
muonic system the 'orbitals' ($d$ and $^8\!B\!e(2^+)$) are of a nuclear
size ($\sim 10^{-13} cm$) and the $d\,^7\!B\!e\,\mu$ molecule has a
size of $\sim 10^{-9} cm$ \cite{bel3}. So the orbitals in our
case are four orders of magnitude smaller than the
molecule and we should expect much more reliable results from
approximation (\ref{lcao}) than in the case of the $H_2^+$ ion.
To continue the analogy with the $H_2^+$ ion we consider the
picture, in which the neutron can be bound in some potential well by
the proton (position a) or almost with the same binding energy by the
$^7\!B\!e$ nucleus (position b), as is shown in Figure 8.

%\vspace{1cm}
%\centerline{\fbox{\rule[-2.5cm]{0cm}{5cm}Insert Figure 8 (called Figure 9 in
%%the FAX)}}
%\centerline{
%{\bf Fig. 8:} a) The neutron is bound by the proton;}
%\centerline{b) it is bound by the $^7\!B\!e$ nucleus.}
%\vspace{1cm}

It is well-known in quantum mechanics that in such systems, where the
possibility exists that a particle can be present in either of the
two potential wells, splitting of levels should occur. This does indeed
happen to a significant extent in this case as we shall now proceed to
demonstrate. From the structure of the channel Hamiltonians $H_{1,2}$
we immediately find for the eigenfunctions $\psi_{1,2}$ that
\begin{equation}
\psi_1 = \Phi_d(\vec r_1) \; \; \Phi_1^{mol}(\vec \rho_1,\vec R_1)
\label{psi1}
\end{equation}
and
\begin{equation}
\psi_2 = \Phi_{^8\!B\!e}(\vec r_2) \; \;
\Phi_2^{mol}(\vec \rho_2,\vec R_2) \; .
\label{psi2}
\end{equation}
Here $\Phi_d$ and $\Phi_{^8\!B\!e}$ are wave functions, which describe
the internal motion in the $d$ and $^8\!B\!e(2^+)$ nuclei, while the
molecular functions for $d\,^7\!B\!e\,\mu$ and $p\,^8\!B\!e\,\mu$ are
given by $\Phi_1^{mol}$ and $\Phi_2^{mol}$, respectively.

Now, after diagonalization of the total four-body Hamiltonian $H$ with
the basis functions (\ref{psi1}) and (\ref{psi2}), we easily find the
coefficients $c_1$ and $c_2$ in equation (\ref{lcao}) and the
transition matrix element $M$,
\begin{equation}
M \sim {\langle\, \Psi \,|\, \psi_2 \,\rangle} \; .
\end{equation}
For the reaction rate $P$ we then have
\begin{equation}
P = \kappa | {\langle\, \Psi \,|\, \psi_2 \,\rangle} |^2 \; ,
\label{tra1}
\end{equation}
where $\kappa$ is the muonic molecule frequency. It is easy to see that
we obtain two solutions of equation (\ref{schr}), $\Psi^{\pm}$,
corresponding to the split levels $E^{\pm}$,
\begin{equation}
\Psi^{\pm} = N^{\pm} ( \lambda^{\pm} \psi_1 + \psi_2 ) \; ,
\end{equation}
where $N^{\pm}$ is the normalization constant and
\begin{equation}
\lambda^{\pm} = \frac{E^{\pm}I-H_{12}}{H_{11}-E^{\pm}} \; ,
\label{lam}
\end{equation}
with the matrix elements
\begin{equation}
I = {\langle\, \psi_1 \,|\, \psi_2 \,\rangle}
\mbox{\hspace{1.5cm}and\hspace{1.5cm}}
H_{ij} = \langle\,\psi_i\,|\,H\,|\,\psi_j\,\rangle \; ,
\end{equation}
which are indeed multidimensional integrals. For the transition rate
(\ref{tra1}) we obtain
\begin{equation}
P^{\pm} = \kappa \frac{|1+I^* \lambda^{\pm}|^2}
{1 + 2Re(I^* \lambda^{\pm}) + |\lambda^{\pm}|^2} \; .
\label{tra2}
\end{equation}
{}From the structure of the channel functions (\ref{psi1}) and
(\ref{psi2}) it is easy to see that for the overlap integral $I$ we
have the relation
\begin{equation}
I = {\langle\, \psi_1 \,|\, \psi_2 \,\rangle}
 \sim \int d\vec \rho \; \; |\Phi^{mol}(\vec \rho,R=0)|^2 \;\;
= \;S_3 \; ,
\label{lap}
\end{equation}
where the value of $S_3$ can be interpreted as a probability for the
heavy particles in muonic molecules to be at small distances
(zero distance in this case). It is easy to see that the matrix
elements of the total four-body Hamiltonian (\ref{fullh}) also
can be expressed through the overlap integral (\ref{lap}). For
example for $H_{11}$ we have
\begin{equation}
H_{11} \approx \varepsilon_1 + S_3 ( V_{n\,B\!e}^{11} +
V_{p\,B\!e}^{11} ) \; ,
\end{equation}
where the $V_{NB\!e}^{11}$ are the matrix elements of the strong
$NB\!e$ potentials over the wave function of the deuteron $\Phi_d$.

Since the nuclear $V_{NB\!e}$ potentials have been determined by
experimentall data \cite{rob}, the most uncertain value, defining the
rate (\ref{tra2}) of the nuclear transition, is $S_3$, the probability
for the heavy particles to be close to each other. To make a rough
estimation we will proceed in the following way. On a qualitative
level the three-body muonic states like $p\,B\!e\,\mu$ can be considered
as two-body $(p+B\!e)$ systems, moving in an effective potential
produced by the muon.
In that potential we can calculate the probability for the heavy
particles to be at small distances. Let us call this value $S_2$.
Then we can suppose that
\begin{equation}
S_3 \approx S_2 = |\Phi (R=0)|^2 \; ,
\end{equation}
The value of the wave function at small distances can be estimated
\cite{com} as
\begin{equation}
|\Phi(R=0)|^2 \leq (m \; \Delta E)^{3/2} \; .
\label{ineq}
\end{equation}
Here $m$ is the reduced mass of the two particles and $\Delta E$ the
separation between the $s$ and $p$ levels in the effective potential,
which produces the wave function $\Phi(R)$. Using this estimation for
the rate of the nuclear transition $P^-$, the only one available in
the process $d\,^7\!B\!e\,\mu \longrightarrow p\,^8\!B\!e\,\mu$, one
obtains
\cite{bel4}
\begin{equation}
P^- = 4.6 * 10^{11} sec^{-1} .
\end{equation}
The inequality (\ref{ineq}) probably overestimates the real value,
since the repulsion in the effective potential is not fully
specified in Reference \cite{com}.

%%%%%%%%%%%%%%%%%%%%%%%%%%%%%%%%%%%%%%%%%%%%%%%%%%%%%%%%%%%%%%%%%%%%%%%%
\bigskip

Now let us come to the description of the nuclear transition in the
molecule $p\,^{10}\!B\!e\,\mu$ \cite{bel5}
\begin{equation}
p \, ^{10}\!B\!e \, \mu \longrightarrow t \, ^8\!B\!e \, \mu + Q \; ,
\mbox{\hspace{2.5cm}} Q \approx 4 keV \; .
\end{equation}
The incredibly small energy release in this transition is related
to the peculiar properties of the $^{10}\!B\!e$ nucleus. Indeed, as one
can see from Figure 9, a double coincidence takes place here: the
coincidence of the position of the excited state $8.4774 MeV$ with
the threshold energy in the channel $^8\!B\!e+2n$, and a coincidence
of the absolute value of the energy of this excited state with the
binding energy of the triton ($8.48MeV$).

%\vspace{1cm}
%\centerline{\fbox{\rule[-2.5cm]{0cm}{5cm}Insert Figure 9 (called Figure 10 in
%%the FAX)}}
%\centerline{
%{\bf Fig. 9:} The spectrum of the $^{10}\!B\!e$ nucleus (schematically)
%\cite{ajz1}.}
%\vspace{1cm}

An additional peculiarity of that transition, in comparison with the
previous one, can be seen from the spectrum of the $^{11}\!B$ nucleus.
It turns out that this nucleus has excited states with energies
which coincide with the threshold energy in the channel
$p+\,^{10}\!B\!e$.
Keeping this in mind, two essentially different mechanisms of the
transition can be formulated, as shown in Figures 10 and 11.

%\vspace{1cm}
%\centerline{\fbox{\rule[-2.5cm]{0cm}{5cm}Insert Figure 10 (called Figure 11 in
%%the FAX)}}
%\centerline{
%{\bf Fig. 10:} Transition through the transfer of two neutrons.}
%\vspace{1cm}
%\centerline{\fbox{\rule[-2.5cm]{0cm}{5cm}Insert Figure 11 (called Figure 12 in
%%the FAX)}}
%\centerline{
%{\bf Fig. 11:} Transition through the resonance state of the
%$^{11}\!B^*$ nucleus.}
%\vspace{1cm}

However, since the spins of the corresponding excited states of the
nucleus $^{11}\!B$ are rather high ($9/2^+$,$5/2^-$), we will neglect
contributions from that mechanism. In other words, our transition
amplitude does not contain the resonant term from Figure 11, due
to a sufficiently small angular momentum in the initial state.
So, we can proceed in complete analogy with the previous case. The
only complications which arise, are due to the more complex wave
function of the nuclear state. To describe the process in the frame of
the 'LCAO' ansatz, we have used a four particle ($2\alpha+2n$) model
for the description of the ground state of the $^{10}\!B\!e$ nucleus,
and a $2\alpha$ model for $^8\!B\!e(g.s.)$.

All we need now are $N\alpha$ and $NN$ potentials, wave functions of
the nuclei $^{10}\!B\!e(g.s.)$, $^8\!B\!e(g.s.)$ and the triton and
molecular wave functions at small distances. The nuclear wave
functions were written in a simple form, reproducing only the
periphery of the nuclei \cite{irv,tan}. For the $N\alpha$ potentials
three different but phaseshift equivalent potentials \cite{sat,dub,dan}
were applied. The results are sensitive to the differences in their
shape. The molecular wave function was constructed by means of the
potential
\begin{equation}
V(R) = 4.5 \,H \, [\frac{2}{\pi}\arctan(aR)-1] \, e^{-bR} - 8 \, H \; ,
\label{pot}
\end{equation}
with parameters fitted to the eigenpotential calculated by means of
the method of surface functions \cite{bel3}. Here,
$H=me^4/\hbar^2=5626.5 eV$ is the Hartree energy unit for the
muon. The potential (\ref{pot}), of course, produces a more reliable
wave function of the molecule at small distances as compared to the
estimation of equation (\ref{ineq}). As $NN$ potentials the
Malfliet-Tjon potentials \cite{mal} have been used.
The results for the reaction rates are shown in Table V.
The fact that the lowest reaction rate occurs with the $N\alpha$
potential given in Reference \cite{dan}, most likely is a reflection
of the repulsive character of that potential.

%%%%%%%%%%%%%%%%%%%%%%%%%%%%%%%%%%%%%%%%%%%%%%%%%%%%%%%%%%%%%%%%%%%%%%%%
\section{New cycling processes and other speculations}

Considering the properties of molecular systems with isotopes of $L\!i$
nuclei, we can hope that, in principle, new cycles analogous to the
one occuring in the $dt$ mixture, can take place. To see this analogy
more clearly, let us remind ourselves of the picture of the cycle
with the $dt\mu$ molecule, as shown in Figure 12.

%\vspace{1cm}
%\centerline{\fbox{\rule[-2.5cm]{0cm}{5cm}Insert Figure 12 (not contained in
%%the FAX) }}
%\centerline{
%{\bf Fig. 12:} The basic reactions of the muon catalyzed $dt$ fusion
%cycle.}
%\vspace{1cm}

Roughly speaking, the two following parameters of that cycle are
specific only for this system: 1) the probability of the resonant
formation of the $dt\mu$ excited state\footnote{Resonance molecular
formation, as it is
well-known \cite{dzh}, also takes place in $d\mu + D_2$ collisions.},
2) the probability of a nuclear transition in the $dt\mu$ molecule
which also has a resonance behavior. As it was pointed out in the
introduction,
the possibility of resonance formation of $d\,L\!i\,\mu$ systems
cannot be
excluded in collisions of the type $d\mu + L\!iD$. If
excited states in $d\,L\!i\,\mu$ systems within the energy range
$E \leq 2.5 eV$ exist, the Vesman mechanism can be invoked.
($E \approx 2.5 eV$ is the dissociation energy of the molecule $L\!iD$
from the $A'\Sigma^+$ state \cite{stw}).

Considering the nuclear resonances we see for both isotopes of
$L\!i$ a situation, which might be even more favourable for nuclear
transitions than in the $dt\mu$ case. Indeed, let us look on the
spectra of the nuclei $^8\!B\!e$ and $^9\!B\!e$, schematically shown in
Figures 13 and 14, respectively.

%\vspace{1cm}
%\centerline{\fbox{\rule[-2.5cm]{0cm}{5cm}Insert Figure 13 (called Figure 14 in
%%the FAX) }}
%\centerline{
%{\bf Fig. 13:} The excited state of $^8\!B\!e$ and the position}
%\centerline{of the threshold for the decay into $d+\,^6\!L\!i\,$
%\cite{ajz1}.}
%\vspace{1cm}
%\centerline{\fbox{\rule[-2.5cm]{0cm}{5cm}Insert Figure 14 (called Figure 15 in
%%the FAX) }}
%\centerline{
%{\bf Fig. 14:} The excited state of $^9\!B\!e$ and the position}
%\centerline{of the threshold for the decay into $d+\,^7\!L\!i$
%\cite{ajz1}.}
%\vspace{1cm}

We see that in both cases there are threshold resonant states
in the compound nuclei $^8\!B\!e$ and $^9\!B\!e$ and the positions of
these
resonances are closer to the corresponding thresholds than the
position of the $^5\!H\!e(3/2^+)$ resonance to the $d+t$ threshold.
In the last case the 'center' (on the energy scale) of the resonance
state $^5\!H\!e(3/2^+)$ is $50 keV$ higher than the threshold energy.
A nuclear transition, therefore, only takes place through the tail
of the resonance curve. Keeping in mind the above pecularities of the
molecular and nuclear structures of the systems $d\mu+\,^{6,7}\!L\!i$,
we can imagine the cycle in the $d+L\!i$ mixture, shown in Figure 15.

%\vspace{1cm}
%\centerline{\fbox{\rule[-2.5cm]{0cm}{5cm}Insert Figure 15 (not contained in
%%the FAX) }}
%\centerline{
%{\bf Fig. 15:}  The basic reactions of the muon catalyzed
%$d\;^{6,7}\!L\!i$ fusion cycle.}
%\vspace{1cm}

The third parameter crucial for the occurence of the cycling process
is the sticking coefficient for the muon in the final state.
Unfortunately, none of these three parameters have till now reliably
estimated.

\bigskip

Other interesting phenomena can be observed in the above systems. We
are talking about the so-called Zeldovich phenomenon or rearrangement
of the spectra of Coulomb systems by a short range
interaction. As it was shown by Zeldovich \cite{zel}, in systems with
long and short range interactions a rearrangement
of the spectrum of the purely long range potential takes place, if the
short range interaction has a resonance or bound state near zero
energy. In this case the strong interaction cannot be treated
perturbatively despite its short range character. Both muonic systems
$d\,^6\!L\!i\,\mu$ and $d\,^7\!L\!i\,\mu$ are indeed good candidates
for the
manifestation of the Zeldovich phenomenon, since the short range
nuclear interactions must generate a resonance behavior at zero
relative energy of $d$ and $L\!i$. It is necessary to emphasize that the
rearrangement of the spectra takes place due to the large size of the
hadronic subsystem at the conditions we have described above.

Considering the nonperturbative situation concerning the influence
of the strong interaction, we can look for the possibility of
nuclear transitions in the corresponding electronic molecules.
As an example let us consider a molecule of ordinary light water,
$H_2\,^{16}\!O$. An interesting property of that molecule is the
coincidence of its energy with the energy of the excited state of the
nucleus $^{18}\!N\!e$. Let us look at the spectrum of this
nucleus, schematically shown in Figure 16.

%\vspace{1cm}
%\centerline{\fbox{\rule[-2.5cm]{0cm}{5cm}Insert Figure 16 (called Figure 17 in
%%the FAX) }}
%\centerline{
%{\bf Fig. 16:} The excited state of $^{18}\!N\!e$ and the position}
%\centerline{of the threshold for the decay into $^{16}\!O+2p$
%\cite{ajz2}.}
%\vspace{1cm}

As one can see, the nucleus $^{18}\!N\!e$ indeed has a state with an
energy
incredibly close (within the accuracy of all measurable figures
\cite{ner}) to the treshold energy for the three-body channel
$^{16}\!O+2p$. Since the binding energy of water is only a few $eV$,
we claim that the water molecule in the rotational state $1^-$ and
the nucleus $^{18}\!N\!e$ in the excited state ($1^-$) of $4.522MeV$
are degenerate states of the same Hamiltonian,
describing 18 nucleons and 10 electrons. This implies that the
wave function of molecular water is not a pure state, but always
contains an admixture of the nuclear state. Since this excited state
of the $^{18}\!N\!e$ nucleus can decay to other channels,
for example, in the channel $^{17}\!F+p$, molecular water should
slowly disappear through the resonance state of the nucleus
$^{18}\!N\!e$.
In other words, the smouldering (slow burning) of water
should take place with an intensity defined by the overlap of the
molecular wave function and the nondecreasing wave function of the
resonance nuclear state. Exactly the same is valid for the molecules of
hydrides of $Li$ due to the coincidence of the nuclear resonance
and threshold energies.

%%%%%%%%%%%%%%%%%%%%%%%%%%%%%%%%%%%%%%%%%%%%%%%%%%%%%%%%%%%%%%%%%%%%%%%%
\section{Conclusions and Outlook}

In this paper we reviewed work performed by our group during the
last few years, devoted to charge nonsymmetric muonic molecules.
For their treatment we have to consider both Coulombic and nuclear
few-body problems. We have given special attention to examples where
the interplay between the Coulomb forces and the nuclear forces
is of atmost importance. We have performed calculations for some
of these examples which have been reviewed here and others will be
published elsewhere.

For the treatment of the purely Coulombic systems, we employed the
method of hyperspherical surface functions. We have calculated
eigenpotentials and eigenfunctions for all isotopes of hydrogen and
$Li$ and some isotopes of $Be$. Using the extreme adiabatic
approximation, our calculations yield no loosely bound states
($\sim 1$ or $2 eV$)
having total angular momentum $L=0$. The nuclear transitions were
calculated in a scheme motivated by the LCAO method, but improved
by taking into account all degrees of freedom of light and heavy
particles. To reduce the numerical effort to manageable proportions
we introduced effective potentials between the nuclear clusters.
Obviously, there is a degree of nonuniqueness in this procedure.
However the main features entering into our calculations, like the
positions of the thresholds, are not effected by this approximation.

We have considered nuclear transitions in the molecules
$d\,^7\!B\!e\,\mu$ and $p\,^{10}\!B\!e\,\mu$, where the energy
releases as well as the momentum
transfers are extremely small on the nuclear scale. For these reasons
the nuclear interactions are expected to have a long range character,
which is important in muonic systems.

We now briefly discuss the reliability of the nuclear reaction rates
performed in these two cases. The crucial ingredient in these
calculations is the value of the molecular wave function at the
origin. In the first instance we calculated this quantity by using
a rather crude approximation, where the Coulomb interactions were
only taken into account in an indirect way suggested by Ref.
\cite{com}. In the second example the Coulomb repulsion at short
distances was taken into account rather reliably, since the
eigenpotentials obtained via the hyperspherical surface function
method have the correct behavior near the origin.

To proceed further we have to keep in mind the difficulties associated
with the Coulombic three-body problem in which we are interested.
In the first place we should emphasize that the systems of the type
$hZ\mu$ with $Z\geq 3$ are highly excited resonances and up to now
no complete solution of neither the Schr\"odinger equation nor the
Faddeev equations has been presented anywhere. The difficulties of
the solution of the Schr\"odinger equation are related mainly to the
very complicated boundary conditions in the configuration space.
Faddeev equations for Coulomb forces are not of Fredholm type in this
range of energies and should therefore be modified to obtain unique
solutions. Additional difficulties arise from the fact that the
calculations of the eigenvalues must be performed to a very high
accuracy. As indicated above, the dissociation energies of the
molecule $L\!iD$ lie around $2.5eV$. That means that the accuracy
of the calculated binding energies of the systems $d\,L\!i\,\mu$ should
be of the order of $1eV$ or even better.

It seems that at the present time only two approaches are
sufficiently promising for the treatment of such systems. The first
one is the hyperspherical surface function expansion, briefly
presented here. From physical reasons we consider resonance states
in the extreme adiabatic approximation as real bound states having
no width. Indeed, the inclusion of the coupling between the channels
corresponds to taking very high excitation energies into
account. The  width appears only, when nonadiabaticity is taken
into account and a coupled system of equations is employed.

The second promising method, which is very often used in molecular
physics \cite{sie}, also is based on the treatment of the resonance
state as a bound state. But the realization of that idea is more
motivated by mathematical arguments. To get a square integrable
solution in this approach, we use coordinate rotation in the complex
plane. It can be shown \cite{sie} that the rotated Hamiltonian has
the same eigenvalues as the physical one.

Concerning the study of the nuclear interaction in the $eV-keV$ energy
region, one can say that it is the only region of energies in the
field of nuclear physics, where there are practically no data
available. We emphasize that the research proposed here can shed light
on the properties and behavior of nuclear systems and interactions
in the very far periphery at distances, exceeding the usual range
of the nuclear interaction by several orders of magnitude, if one
considers muonic systems.

%%%%%%%%%%%%%%%%%%%%%%%%%%%%%%%%%%%%%%%%%%%%%%%%%%%%%%%%%%%%%%%%%%%%%%%%
\bigskip
{\bf Acknowledgment. }
One of the authors (V.B.B.) would like to thank the Physikalisches
Institut der Universit\"at Bonn for its hospitality and two of them
(V.B.B. and W.S.) would like to acknowledge financial support by the
Scientific Division of NATO, Research Grant No CRG 930102.

%%%%%%%%%%%%%%%%%%%%%%%%%%%%%%%%%%%%%%%%%%%%%%%%%%%%%%%%%%%%%%%%%%%%%%%%
\vspace{1cm}
\begin{table}
\caption{Binding energies of ($dt\mu$) with $L=0$}
\begin{tabular}{c|cc}
state       & ground state & excited state \\ \hline
present     & -295.54      & -35.80 \\
\cite{mon} & -319.14      & -34.83
\end{tabular}
\end{table}
\begin{table}
\caption{Binding energies of ($h\,^6\!L\!i\,\mu$) with $L=0$}
\begin{tabular}{c|ccc}
system      &  $p\,^6\!L\!i\,\mu$ & $d\,^6\!L\!i\,\mu$ &
$t\,^6\!L\!i\,\mu$ \\ \hline
present     & -24.3       & -23.8      & -35.3       \\
\cite{kra1} & -17.6       & -18.5      & -19.8
\end{tabular}
\end{table}
\begin{table}
\caption{Binding energies of ($h\,^7\!L\!i\,\mu$) with $L=0$}
\begin{tabular}{c|ccc}
system      &  $p\,^7\!L\!i\,\mu$ & $d\,^7\!L\!i\,\mu$ &
$t\,^7\!L\!i\,\mu$ \\ \hline
present     & -20.8       & -25.9      & -37.5       \\
\cite{kra1} & -21.0       & -22.0      & -23.3
\end{tabular}
\end{table}
\begin{table}
\caption{Binding energies of ($h\,^7\!B\!e\,\mu$) with $L=0$}
\begin{tabular}{c|cc}
system      &  $p\,^7\!B\!e\,\mu$ & $d\,^7\!B\!e\,\mu$ \\ \hline
present     & -11.7      & -29.3
\end{tabular}
\end{table}

%\vspace{0.5cm}
\centerline{
TABLE V. Reaction rates for the transition $p\,^{10}\!B\!e\,\mu
\longrightarrow t\,^8\!B\!e\,\mu$}
\centerline{\hspace{1cm}calculated with the three different
$\alpha N$ potentials \cite{sat,dub,dan}.}
\vspace{-0.5cm}
\begin{table}
\begin{tabular}{cccc}
Potential        & \cite{sat} & \cite{dub} & \cite{dan} \\ \hline
$P^- [sec^{-1}]$ & 1688       & 1129       & 238
\end{tabular}
%\caption{Reaction rates for the transition $p\, ^{10}\!B\!e\,\mu
%\longrightarrow t\; ^8\!B\!e\,\mu$ calculated with the three different
%$\alpha N$ potentials \cite{sat,dub,dan}}
\end{table}

{\bf Figure Captions.}\\
FIG.1. The correlation between the abundance of $L\!i$ and $F\!e$ in
       old stars (normalized to the abundance in the sun).\\
FIG.2. The correlation between the abundance of $M\!g$ and $F\!e$ in
       old stars (normalized to the abundance in the sun).\\
FIG.3. $Y_P$ denotes the predicted primordial abundance of $^4\!H\!e$
       relative to $H$ as a function of $\eta$ for three different
       numbers of neutrino species $N_\nu$. The other curves show the
       relative abundances of $D$, $^3\!H\!e$, and $^7\!L\!i$ as a
       function of $\eta$.\\
FIG.4. New experimentall data (at energies less than $100 keV$) and old
       data (above $100 keV$) for the S-factor. The full calculation
       (solid line) is shown together with a calculation using only the
       ($L=2$,$S=2$) configuration (short-dashed line).\\
FIG.5. Jacobi coordinates for the system $hZ\mu$.\\
FIG.6. The spectrum of the $^8\!B\!e$ nucleus (schematically)
       \cite{ajz1}.\\
FIG.7. Two different sets of Jacobi coordinates appropriate to describe
       the initial and the final states of the four-body system.\\
FIG.8. a) The neutron is bound by the proton
       b) it is bound by the $^7\!B\!e$ nucleus.\\
FIG.9. The spectrum of the $^{10}\!B\!e$ nucleus (schematically)
       \cite{ajz1}.\\
FIG.10. Transition through the transfer of two neutrons.\\
FIG.11. Transition through the resonance state of the $^{11}\!B^*$
        nucleus.\\
FIG.12. The basic reactions of the muon catalyzed $dt$ fusion cycle.\\
FIG.13. The excited state of $^8\!B\!e$ and the position of the
        threshold for the decay into $d+\,^6\!L\!i\,$ \cite{ajz1}.\\
FIG.14. The excited state of $^9\!B\!e$ and the position of the
        threshold for the decay into $d+\,^7\!L\!i$ \cite{ajz1}.\\
FIG.15. The basic reactions of the muon catalyzed $d\;^{6,7}\!L\!i$
        fusion cycle.\\
FIG.16. The excited state of $^{18}\!N\!e$ and the position of the
        threshold for the decay into $^{16}\!O+2p$ \cite{ajz2}.\\

%%%%%%%%%%%%%%%%%%%%%%%%%%%%%%%%%%%%%%%%%%%%%%%%%%%%%%%%%%%%%%%%%%%%%%%%

\end{document}